\documentclass[aps, twocolumn, reprint, amsmath, amssymb]{revtex4-1}
\usepackage[utf8]{inputenc}
\usepackage{graphicx}
\usepackage{hyperref}
\usepackage{bm}

\newcommand{\uE}{\ensuremath{\bm{u}_\mathrm{E}}}
\newcommand{\ud}{\ensuremath{\bm{u}_\mathrm{de}}}
\newcommand{\curv}[1]{\ensuremath{\mathcal{K}\left( #1 \right)}}
\newcommand{\ntilde}{\ensuremath{\widetilde{n}}}

\newcommand{\Cs}{\ensuremath{C_\mathrm{s}}}
\newcommand{\Te}{\ensuremath{T_\mathrm{e}}}
\newcommand{\mi}{\ensuremath{m_\mathrm{i}}}

\newcommand{\geff}{\ensuremath{g}}

\newcommand{\Omegaci}{\ensuremath{\Omega_\mathrm{ci}}}

\newcommand{\tmax}{\ensuremath{t_{\max V}}}
\newcommand{\deltantrans}{\ensuremath{\left( \triangle n / N \right)_\mathrm{c}}}

\newcommand{\K}[1]{\ensuremath{\mathcal{K} \left( #1 \right)}}

\newcommand{\Ref}[1]{[\onlinecite{#1}]}
\newcommand{\Eqnref}[1]{Eq.\ \ref{eq:#1}}
\newcommand{\Eqsref}[1]{Eqs.\ \ref{eq:#1}}
\newcommand{\Figref}[1]{Fig.\ \ref{fig:#1}}
\newcommand{\Figsref}[1]{Figs.\ \ref{fig:#1}}

\begin{document}
\title{Amplitude and size scaling for interchange motions of plasma filaments}
\author{R.\ Kube}
\email[E-mail: ]{ralph.kube@uit.no}
\affiliation{Department of Physics and Technology, UiT The Arctic University of Norway, N-9037 Tromsø, Norway}
\author{M.\ Wiesenberger}
\affiliation{Institute for Ion Physics and Applied Physics, 
                     Universität Innsbruck, A-6020 Innsbruck, Austria}
\author{O.\ E.\ Garcia}
\affiliation{Department of Physics and Technology, UiT The Arctic University of Norway, N-9037 Tromsø, Norway}

\date{\today}
\begin{abstract}
The interchange dynamics and velocity scaling of blob-like plasma filaments are
investigated using a two-field reduced fluid model. For incompressible flows due to
buoyancy the maximum velocity is proportional to the square root of the relative
amplitude and the square root of its cross-field size. For compressible flows in a
non-uniform magnetic field this square root scaling only holds for ratios of
amplitudes to cross-field sizes above a certain threshold value. For small amplitudes
and large sizes, the maximum velocity is proportional to the filament amplitude. The
acceleration is proportional to the amplitude and independent of the cross-field size
in all regimes. This is demonstrated by means of numerical simulations and explained
by the energy integrals satisfied by the model.
\end{abstract}

\maketitle
\section{Introduction}
At the outboard mid-plane of magnetically confined plasmas one universally observes radial
motion of field-aligned plasma pressure perturbations. These are structures of excess
pressure localized in the plane perpendicular to the magnetic field and are therefore
commonly referred to as blobs \Ref{zweben-1985, antar-2001}.  These blobs are
intermittently created close to the last closed magnetic flux surface 
\Ref{boedo-2003, terry-2003, agostini-2007, garcia-2013} 
and propagate radially outwards through the scrape-off layer, mediating a significant loss
channel for particles and heat. They may further be responsible for high rates of plasma
particle recycling at the main chamber wall \Ref{umansky-1998, lipschultz-2001}. 
Experimental studies suggest that their dynamical properties set the profile length 
scale of the particle density profile in the scrape-off layer 
\Ref{labombard-2001, carralero-2015, kube-2016}. 
A stochastic model, which models density fluctuation time series
in the scrape-off layer as the superposition of exponentially decaying pulses,
explicitly relates the scale length of the density profile to the average
radial blob velocity \Ref{garcia-2016}.

A large body of research suggests that interchange motions due to the non-uniform
magnetic field is the mechanism underlying blob propagation 
\Ref{krash-2001, dippolito-2002, bian-2003, dippolito-2004, garcia-2005, dippolito-2011}.
At the outboard mid plane of a magnetically confined plasma magnetic gradient and
curvature drifts guide electrons and ions in opposing directions. As a consequence, a blob
of excess pressure will be electrically polarized, generating a dipolar potential
structure that is out of phase with the pressure perturbation. The resulting electric
drift propagates the plasma blob radially outwards, away from the closed field line
region, thereby exchanging hot and dense with cold, low density plasma 
\Ref{krash-2001, dippolito-2004, garcia-2006, garcia-2006-ps, dippolito-2011}.
Analysis of data time series shows that the observed motion of the plasma blobs agrees
well with the suggested theory 
\Ref{boedo-2003, rudakov-2005, grulke-2006, grulke-2014, kube-2016, theodorsen-2016}.

Plasma filaments have also been investigated in basic toroidal plasma
experiments.  In cold plasma experiments performed at the Versatile Toroidal Facility it
was observed that plasma blobs develop a mushroom shape, as often observed in numerical
simulations, and that their flow field is dipolar \Ref{katz-2008}.  Experiments performed
in an open field line configuration at the TORPEX device further corroborate that the
interchange mechanism supports blob propagation \Ref{theiler-2009, furno-2011, riva-2016}. 

In the equatorial F-layer ionosphere, so-called equatorial spread-F plasma
depletions, or bubbles, have been observed to propagate radially out across the boundary
of the F-layer ionosphere \Ref{hysell-2000}. Recent measurements of magnetospherical
plasmas at Saturn \Ref{rymer-2009} and analytic theory \Ref{wolf-2012} suggest that plasma
bubbles propagating through plasma sheets in these regions are also driven by buoyancy.

The simplest fluid models used to describe the interchange dynamics of seeded plasma blobs
feature the advection of the particle density by the electric drift \Ref{krash-2001,
dippolito-2003, myra-2004, garcia-2005}.  For self-consistent dynamics the electric
potential is computed by invoking quasi-neutrality in the low-frequency limit, often
applying the so-called Boussinesq approximation. Within this approximation, which is valid
for small particle density perturbations $\widetilde{n}$ relative to the background $N$,
$|\widetilde{n}| / N \ll 1$, one neglects particle density gradients in the inertial terms
of the fluid equations while retaining them in the other terms. Recent work avoids
this simplification \Ref{yu-2006, madsen-2011, wiesenberger-2014, angus-2014, omotani-2015, held-2016}.

In previous works a scale analysis of the model equations was employed to derive velocity
scaling laws for blobs, \Ref{myra-2005, garcia-2005, theiler-2009, kube-2011, manz-2013}.
In the simplest case only particle density fluctuations on a uniform background are considered
and the parallel dynamics are neglected. Scale analysis of this model suggests that the
radial center of mass velocity of blobs follows the so-called \emph{inertial} scaling 
$V \sim \left(\ell \triangle n / R_0 N \right)^{1/2}$ \Ref{garcia-2005}. Here, $\ell$ is
the initial blob cross-field size, $R_0$ is the major radius, and $\triangle n$ the
initial blob amplitude. 

Numerical simulations of incompressible fluid models recover the inertial velocity 
scaling in the case of small initial blob amplitudes \Ref{kube-2011}.  Numerical
simulations relaxing the Boussinesq approximation further validate the inertial
velocity scaling for blobs with moderate initial amplitudes \Ref{angus-2014,
omotani-2015, held-2016}.  First indications that the inertial velocity scaling
does not hold for blob motions with small amplitudes using a model with
compressible flows were reported in \Ref{wiesenberger-2014, olsen-2016}.

In this contribution we review the simplifications leading to the reduced two-field fluid
models and discuss the common practice of neglecting drift compression terms in the
particle continuity equation. This is often justified due to the smallness of the terms
and has been applied throughout the literature
\Ref{bian-2003, dippolito-2003, myra-2004, garcia-2005, myra-2006, yu-2006, garcia-2006, krash-2008, kube-2011, angus-2012, angus-2014}.
We derive conservation laws from the model equations and discuss how neglecting drift
compression terms impacts the conservation properties of the model equations. Velocity
scaling laws are derived from the conservation laws and compared to the velocity scaling
usually derived from the vorticity equation.  Numerical simulations of seeded blob
propagation are performed with and without drift compression terms in the continuity
equation, using two different sets of numerical methods. The resulting velocity scaling of
the seeded plasma blobs is compared to analytically derived scaling laws.

\section{Analytic modeling}
For an isothermal, quasi-neutral plasma with a single cold ion species, the low-frequency
electrostatic dynamics is described by the continuity equation for the plasma particle
density $n$,

\begin{align}
    \frac{\partial n}{\partial t} + \nabla \cdot \left( n \uE +  n \ud \right)  & = \nu \nabla_\perp^2 n.
\end{align}
Here, the electric drift is given by $\uE = \bm{b} \times \nabla \phi / B$ and the
diamagnetic drift by $\ud = -(\Te / e n B) \bm{b} \times \nabla n$, where $\phi$ is the
electric potential, $\bm{B} = B \bm{b}$ the magnetic field, $e$ the elementary charge and
$\Te$ the electron temperature. We consider a slab magnetic field given by 
$\bm{B} = (B_0 R_0 / x) \bm{e}_z$ in a Cartesian coordinate system.  This field
approximates the magnetic field at the outboard mid plane using the radial coordinate $x$,
the approximately poloidal coordinate $y$ and the $z$ direction aligned to the magnetic
field.  We further assume that the aspect ratio $R_0 / a$, where $R_0$ is the major and
$a$ is the minor radius, is small as to approximate $1 / B \approx 1 / B_0$ within the
bounds of the model.  In this approximation the curvature of the magnetic field vanishes, 
$\left( \bm{b} \cdot \nabla \right) \bm{b} = \bm{0}$. Introducing the operator
$\K{u} \equiv \nabla\cdot\left( {\bm{b}\times\nabla u}/{B} \right) = -(1/B_0 R_0)\partial u / \partial y$
allows us to write the drift compression terms as $\nabla \cdot \left( n \ud \right) = (\Te / e) \K{n}$ and
$\nabla \cdot \uE = -\K{\phi}$, respectively. 

The particle density is now separated into a stationary and homogeneous background $N$ and
a perturbation $\widetilde{n}$ as $n = N + \widetilde{n}$, where we assume that the
relative perturbation amplitude is small, $|\widetilde{n}| / N \ll 1$.  An energetically
consistent model that describes the dynamics of density perturbations in the limit of the
Boussinesq approximation is then given by
\begin{subequations}
\begin{align}
    \left( \frac{\partial}{\partial t} + \frac{\bm{b} \times \nabla \phi}{B_0} \cdot \nabla \right)& \frac{\ntilde}{N} 
    + \alpha \curv{\phi} - \beta \frac{\Te}{e} \curv{\frac{\ntilde}{N}} = \nu \nabla_\perp^2 \frac{\ntilde}{N}  \label{eq:model_n}\\
    \left( \frac{\partial}{\partial t} + \frac{\bm{b} \times \nabla \phi}{B_0} \cdot \nabla \right)& \Omega
    - \Omegaci \frac{\Te}{e}\curv{\ntilde} = \nu \nabla_\perp^2 \Omega. \label{eq:model_omega}
\end{align}
\label{eq:model_dimensional}
\end{subequations}
Here we have introduced the field-aligned vorticity density of the electric drift
$\Omega = N \nabla^2_\perp\phi/B_0 \approx N \bm{b} \cdot \nabla \times \uE $ and the ion 
cyclotron frequency $\Omegaci = e B_0 / \mi$. Artificial coefficients 
$\alpha, \beta \in \{0,1\}$ in front of the drift compression terms in \Eqnref{model_n}
allow to isolate the contribution of these terms on the blob dynamics. Choosing 
$\alpha = \beta = 0$ describes a plasma in a homogeneous magnetic field experiencing 
a gravitational drift with $\geff = - {\Te}/{(R_0 \mi)} \equiv  -\Cs^2 / R_0$ as the gravity.
This may be used to describe astrophysical plasmas, specifically ionospherical irregularities 
such as equatorial spread F phenomena, which are thought to be caused by the interchange 
instability \Ref{ott-1978, hysell-2000}. The model with $\alpha = \beta = 1$ also arises 
when taking the long wavelength limit of a delta-f gyrofluid model \Ref{wiesenberger-2014} 
and describes compressible electrostatic motions in a non-uniform magnetic field.

To study the evolution of seeded blobs, the density field is initialized as a Gaussian 
function with no initial vorticity as 
\begin{subequations}
\begin{align}
    \tilde n(\bm{x}, t=0) & = \triangle n \exp \left( -\frac{\bm{x}^2}{2\ell^2} \right), \label{eq:init_n}\\
    \Omega(\bm{x}, t = 0) & = 0. \label{eq:init_omega}
\end{align}
\label{eq:init}
\end{subequations}
The initial perturbation amplitude of the blob is given by $\triangle n$ and
$\ell$ is its characteristic cross-field size.
To obtain conservation laws of \Eqnref{model_dimensional} we multiply \Eqnref{model_n}
with $-N \Te x / R_0$ and $\tilde{n} \Te$ respectively and  \Eqnref{model_omega} with 
$-e \phi/\Omega_{ci}$ and integrate the resulting equations over the domain \Ref{garcia-2006-ps}. 
Adding the results yields two conservation laws
\begin{subequations}
    \begin{align}
      \frac{\mathrm{d}}{\mathrm{d} t} \left( G + E \right) &= \Lambda_G - \Lambda_E  , 
    \label{eq:conservation_1}\\
      \frac{\mathrm{d}}{\mathrm{d} t} \left( S + \alpha E \right) &= -\left( \Lambda_S + \alpha \Lambda_E \right),  
    \label{eq:conservation_2}
\end{align}
\label{eq:conservation}
\end{subequations}
with the energy integrals defined as
\begin{subequations}
\begin{align}
    G(t) &:= \mi \geff\int \mathrm dA\, \tilde n x, \label{eq:energy_int_G}\\
    E(t) &:= \frac{1}{2} \mi N \int \mathrm{d}A\, \left(\frac{\nabla_\perp\phi} {B_0}\right)^2 , \label{eq:energy_int_E} \\
    S(t) &:= \frac{1}{2}N \Te \int \mathrm{d}A\, \left( \frac{\tilde n}{N} \right)^2.
    \label{eq:energy_int_S}
\end{align}
\label{eq:energy_integrals}
\end{subequations}
They correspond to the potential energy $G$ of the plasma in its effective gravity field,
the kinetic energy $E$ and an entropy-like quantity $S$. For the partial integration over
the spatial domain we assume boundary terms to vanish. The energy dissipation is
expressed by 
$\Lambda_G := - \nu \int\mathrm{d}A\, \Te x\nabla^2_\perp n / R_0$, 
$\Lambda_E := \nu \int \mathrm{d}A\, \mi \Omega^2/N$ and 
$\Lambda_S := \nu\int\mathrm{d}A\, \Te \left( \nabla_\perp \tilde{n} \right)^2/N$.
Then \Eqnref{conservation_1} may be interpreted as the evolution of the non-linearly
conserved energy of the system while \Eqnref{conservation_2} expresses non-linear
conservation of a free-energy like quantity. With \Eqsref{init}, initial conditions on
\Eqsref{energy_integrals} are given by
$S(0) = N \Te \pi \ell^2 (\triangle n/N)^2 / 2$
and 
$G(0) = E(0) = 0$.

The transfer between potential and kinetic energy, as well as between kinetic energy
and the entropy is mediated by the coupling term
$\mathrm{d} E/{\mathrm{d} t} = -\mathrm{d} G/\mathrm{d} t = \int \mathrm{d}A\, \tilde{n} \Te \mathcal{K}(\phi)$, 
where we neglect diffusion.  It describes a transfer of potential energy of a plasma
structure in an effective gravitational field into kinetic energy. Including compression
of the electric drift, $\alpha = 1$, this coupling term mediates a transfer of $S$ into
the kinetic energy of the system \Ref{scott-2005}. While \Eqnref{conservation_1} sets no
bound on either $G$ or $E$, \Eqnref{conservation_2} is a restriction for $E$ since both
$S\geq 0 $ and $E\geq 0$.  In other words, the compression of the electric drift
introduces an upper bound on the kinetic energy through conservation of internal energy as
described by \Eqsref{conservation}.

To obtain a velocity scaling of seeded plasma blobs we perform an order of magnitude 
estimate by assuming that the dynamics of the flow is due to inertia
$\partial / \partial t \sim \Omega \sim \phi /B \ell^2$, 
as well as $\left(1 / N \right) \partial n / \partial y \sim \triangle n / (N \ell)$
in \Eqnref{model_omega} to obtain \Ref{garcia-2005}

\begin{align}
    \frac{\max V}{\Cs} = \mathcal{R} \left(\frac{\ell }{R_0}\frac{\triangle n}{N} \right)^{1/2}. \label{eq:vorticity_scaling}
\end{align}
Here, $\mathcal{R}$ is a proportionality factor that will later be determined from
numerical simulations. This scaling is valid for both compressible and incompressible
flows and is often called the inertial scaling \Ref{garcia-2005}. In this regime, the
velocity is proportional to the square root of the blobs cross-field size $\ell$ and its
relative perturbation amplitude $\triangle n / N$.

To obtain a velocity scaling law from the conservation laws given by \Eqsref{conservation}
we introduce the radial center of mass coordinate of a plasma blob \Ref{garcia-2005}
\begin{align}
    X(t) & = \frac{\int \mathrm{d}A\, n x}{M},
\end{align}
where $M = \int \mathrm{d}A \tilde n = 2\pi \ell^2\triangle n $ is the conserved mass. 
Using \Eqsref{model_dimensional} the radial center of mass velocity 
$V(t) = \mathrm{d} X / \mathrm{d} t$ can be written as \Ref{garcia-2006}
\begin{align}
    V(t) & = - \frac{1}{M} \int \mathrm{d}A\, \tilde n \frac{1}{B_0}\frac{\partial \phi}{\partial y}. \label{eq:vcom}
\end{align}
The conservation law given by \Eqnref{conservation_2} yields bounds on the entropy and 
energy as $S(t) \leq S(0)$ and $E(t) \leq S(0)$. An upper bound on the center of mass 
velocity given by \Eqnref{vcom} is then found by applying the Cauchy-Schwarz inequality as
\begin{align}
    \left( M V \right)^2 & = \left( \int \mathrm{d}A\, \tilde n \frac{1}{B_0} \frac{\partial \phi}{\partial y} \right)^2 \nonumber \\
                         & \leq \int \mathrm{d}A\, \tilde n^2 \int \mathrm{d}A\, \left(\frac{ \nabla_\perp \phi}{B_0}\right)^2 \nonumber \\
                         & \leq \frac{4}{\mi \Te} S^2(0). \label{eq:energy_cschwartz}
\end{align}
This shows that the center of mass velocity is bounded by the initial entropy of
the plasma. This initial entropy on the other hand is set by \Eqnref{init_n}, such that
a blobs initial amplitude may set an upper limit on its maximal radial velocity.

Using the initial conditions on \Eqsref{energy_integrals} we evaluate this upper
bound on the center of mass velocity as
\begin{align}
  \frac{\max V}{\Cs} &=  \frac{\mathcal{P}}{ 4} \frac{\triangle n}{N}. \label{eq:energy_scaling}
\end{align}
Here $\mathcal{P}$ is a numerical coefficient with $0 < \mathcal P \leq 1$ that will 
later be determined from numerical simulations. This upper bound on the center of 
mass velocity is a direct consequence of energy conservation and must hold at any stage 
of the blobs evolution in the case of a non-uniform magnetic field.

We equate \Eqsref{vorticity_scaling} and \ref{eq:energy_scaling} to evaluate the critical 
ratio of initial amplitude to size, above which the velocity is constrained by 
\Eqnref{vorticity_scaling} rather than the linear scaling given by \Eqnref{energy_scaling}
as
\begin{align}
  \left( \frac{\triangle n/N}{\ell/R_0}\right)_\mathrm{c} = \left(\frac{4 \mathcal R}{\mathcal P}\right)^2. \label{eq:transition}
\end{align}
In the case of a non-uniform magnetic field we thus expect large amplitude blobs with
small cross-field sizes to be subject to the inertial velocity scaling given by 
\Eqnref{vorticity_scaling}.

To find a scaling for the center of mass acceleration we rewrite 
\Eqnref{energy_cschwartz} as
\begin{align}
    \left( M V \right)^2 & \leq \frac{4 S(0)}{\mi \Te} E(t), 
\end{align}
which is true for both the compressible and the incompressible case. Further assuming 
that the blob accelerates uniformly in the initial phase,
\Ref{myra-2004, garcia-2005, garcia-2006, kube-2011, held-2016, wiesenberger-2014},
$V = A t$ and $X = A t^2 / 2$, and using \Eqnref{conservation_1}, the time derivative 
yields
\begin{align}
  A = \frac{\mathcal Q}{2} \frac{\triangle n}{N} \frac{\Cs^2}{R_0} 
    \equiv \frac{\mathcal{Q}}{2} \geff \frac{\triangle n }{N}, 
    \label{eq:acceleration}
\end{align}
where again $\mathcal Q$ is a numerical coefficient with $0<\mathcal Q \leq 1$.
This shows that the blobs is always accelerated with a rate given by the effective gravity 
$\Cs^2 / R_0$ and its initial perturbation amplitude. Such a uniform acceleration is in 
accordance with previous work where a scale analysis suggests that the temporal scale of 
the interchange motions described by \Eqnref{model_omega} is given by 
$\gamma = \left( \geff \triangle n / \ell N \right)^{1/2}$ \Ref{garcia-2005}.

Finally, we introduce the time it takes a blob to accelerate to its maximal velocity
as $\tmax$. Using $\max V = \tmax A$ we evaluate this time to be 

\begin{align}
    \frac{\tmax}{R_0 / \Cs} = \frac{2 \mathcal{R}}{\mathcal{Q}}
    \begin{cases}
        \displaystyle{\frac{\mathcal{P}}{4 \mathcal{R}}}&  \text{for } \displaystyle{\frac{\ell / R_0}{\triangle n / N} > \left(\frac{\mathcal{P}}{4 \mathcal{R}}\right)^2} \\ 
        \displaystyle{\sqrt{\frac{\ell / R_0}{\triangle n / N}}}& \text{otherwise}
    \end{cases}
\label{eq:tmaxv}
\end{align}
Thus, as a consequence of the constant acceleration phase, the time it takes the
blob to achieve its maximal radial velocity is independent of its initial amplitude or size
when the amplitude is small and the size is large. In the opposite regime, large amplitude 
blobs feature a shorter acceleration phase than small amplitude blobs and vice versa for the 
blob size.


\section{Numerical simulations}
We continue by investigating the effect of the drift compression terms
$\nabla \cdot \left( n \ud \right)$ and $n \nabla \cdot \uE $ in \Eqnref{model_n} 
on the center of mass dynamics of seeded plasma blobs by numerical simulations.
To this end we normalize the spatial scales to $\ell$, the temporal scale to the 
interchange rate $\gamma = \left( {\Cs^2 / (R_0 \ell)} \right)^{1/2}$, the electric 
potential as $\phi \rightarrow \widehat{\phi} = \phi / (\gamma B_0 \ell^2)$, and the 
vorticity density as $\Omega \rightarrow \widehat\Omega = \Omega/N\gamma$ to rewrite 
\Eqsref{model_dimensional} in dimensionless form
\begin{subequations}
\begin{align}
    \frac{\partial n}{\partial t} + \{\phi, n\} - \alpha \kappa \frac{\partial \phi}{\partial y} + 
        \beta \delta \frac{\partial n}{\partial y}
        & = \nu \nabla^2_\perp n, \label{eq:reduced_n} \\
    \frac{\partial \Omega}{\partial t} + \{\phi, \Omega\} + \frac{\partial n}{\partial y} 
        &  =  \nu \nabla^2_\perp \Omega. \label{eq:reduced_omega}
\end{align}
\label{eq:model_dimensionless}
\end{subequations}
The free parameters of this model are $\kappa \equiv \ell / R_0$, which sets the
cross-field size of the plasma blob as a fraction of the major radius and
$\delta \equiv \gamma / \Omegaci \equiv \rho_s / \left(R_0 \ell\right)^{1/2}$. This
parameter gives the ratio of the interchange time scale relative to the ion cyclotron frequency,
or alternatively, the ratio of the thermal gyroradius $\rho_s = \left(\Te \mi\right)^{1/2}/ e B_0$ 
to the geometric mean of the blobs cross-field size and the major radius.
The drift advection terms are written using the Poisson bracket formalism 
$\{f, g\} = \partial_x \left(f \partial_y g\right) - \partial_y \left(f \partial_x g\right)$.
Typical scrape-off layer parameters are chosen by setting
$R_0 = 1 \mathrm{m}$, $B_0 = 1 \mathrm{T}$, $\Te = 10 \mathrm{eV}$, and a typical blob 
cross-field size $\ell=1$cm, which yields the dimensionless parameters
$\delta = 4.6 \times 10^{-3}$ and $\kappa = 10^{-2}$. We choose $\nu = 10^{-3}$ such that 
dissipation is much smaller than effective buoyancy \Ref{garcia-2005}. 
The respective simulations are labeled as 
no compression $\alpha = \beta = 0$,
no electric drift compression $\alpha = 0, \beta = 1$,
no diamagnetic drift compression $\alpha = 1, \beta = 0$,
as well as full compression $\alpha = \beta = 1$
throughout the rest of this contribution.

Equations \ref{eq:model_dimensionless} were solved with initial conditions given by
\Eqsref{init} using a spectral Fourier-Galerkin method to discretize spatial 
derivatives \Ref{2dads}, as well as by discontinuous Galerkin methods 
(cf. \emph{FELTOR} library \Ref{wiesenberger-phd}) for comparison. The detailed 
numerical codes including the input parameters as well as all output data can be found 
in the supplemental data to this contribution \Ref{numerics-supplemental}. The results of 
the simulations were tested for convergence by increasing the domain size together with 
the number of cells and discretization points, as well as by reducing the diffusion coefficient, 
until no change in the blob dynamics was observable. The energy equations \Eqsref{conservation} 
were verified numerically and we found negligible differences between the discontinuous and 
the Fourier-Galerkin methods.

\begin{figure}[htb]
    \includegraphics[width=\columnwidth]{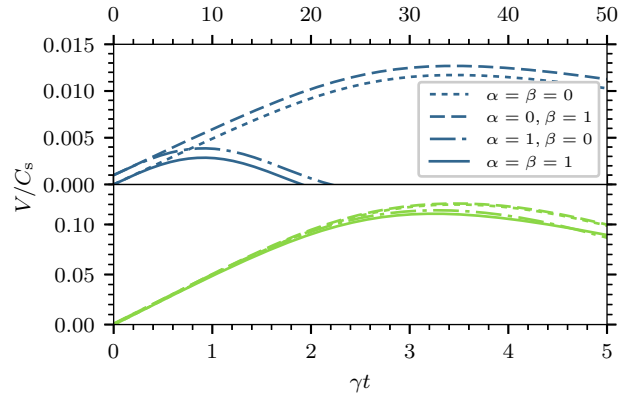}
    \caption{(Color)The radial center of mass velocity of a blob with $\triangle n/N = 0.02$ 
        (upper panel) and $\triangle n / N = 2$ (lower panel). An offset of $10^{-3}$ 
        is added to the dashed and dashed-dotted line for visibility.}
    \label{fig:vcom_t}
\end{figure}

Figure \ref{fig:vcom_t} shows the center of mass velocity \Ref{garcia-2005, wiesenberger-2014} 
of the blob as a function of time for $\triangle n / N = 0.02$ and $\triangle n / N = 2$. 
In the case of a small initial amplitude, $\triangle n / N = 0.02$, the
blobs center of mass velocity initially increases approximately linearly in time for all four
simulated cases. When neglecting electric drift compression, $\alpha = 0$, the
blob assumes a maximal radial velocity, $\max V \approx 1.2 \times 10^{-2} \Cs$
at $\tmax \approx 32 \gamma^{-1}$. Including electric drift compression, $\alpha = 1$,
shortens the period of uniform acceleration. In this case the blob assumes
a maximal radial center of mass velocity of $\max V \approx 3 \times 10^{-3} \Cs$
at $\tmax \approx 10 \gamma^{-1}$. After this initial acceleration phase the blob decelerates
and shows dispersion due to non-linear mixing \Ref{garcia-2005, garcia-2006}.

For $\triangle n / N = 2$ the blob dynamics is independent of the included compressional
terms in the model equations. After an approximately uniform acceleration phase the blob 
assumes a maximal radial velocity of $\max V \approx 0.11 \Cs$ at $\tmax \approx 3 \gamma^{-1}$. 
We conclude that the electric drift compression has a profound influence on the dynamics of 
the blob in the case of small initial blob amplitudes.

\begin{figure}[htb]
    \includegraphics[width=\columnwidth]{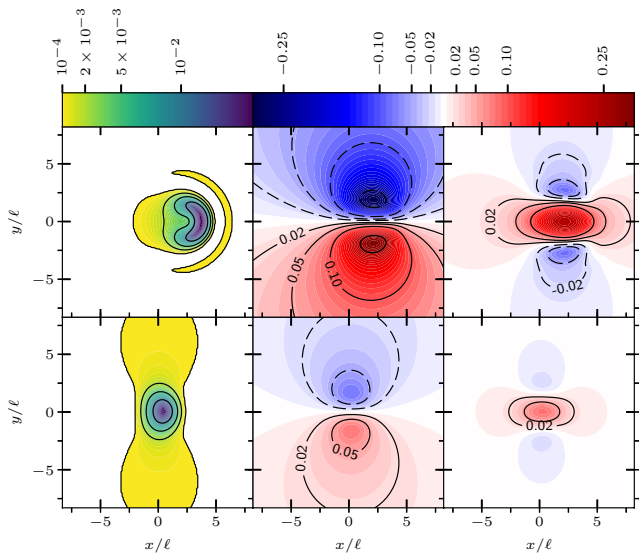}
    \caption{(Color) Contour plots of the particle density perturbation (left column), 
             the electrostatic potential (middle column) and the radial electric drift 
             component (right column) for simulations without drift compression 
             (upper row) and including drift compression terms (lower row). The initial 
             amplitude is given by $\triangle n / N = 0.02$ and the fields are shown at 
             the time each blob propagates approximately at its maximal radial velocity, 
             $t = 35\gamma^{-1}$ in the upper row and $t=10\gamma^{-1}$ in the lower row.
             The black lines denote equi-density and equi-potential surfaces and 
             correspond to the ticks in the colorbars.}
    \label{fig:contour_vmax}
\end{figure}

Physically, the compressibility of the electric drift arises from the inhomogeneity of the
magnetic field. The effect of including this term in the model equations is visualized in
\Figref{contour_vmax}. We show the evolution of a blob with $\triangle n / N = 0.02$
taken from simulations with $\alpha = \beta = 0$ at $t = 10 \gamma^{-1}$ in the upper
column and at $t = 35 \gamma ^{-1}$ for $\alpha = 1, \beta = 0$ in the lower column. These
are the times at which the blobs radial center of mass velocity is approximately maximal.
The left column shows the particle density and the middle column the electric potential,
where equi-potential lines give the flow field on which plasma is advected by the electric
drift.  In the right column we present the radial component of the electric drift,
$-\partial \phi / \partial y$, which corresponds to the compression of the electric drift
via $\nabla \cdot \uE = -\left( \partial \phi / \partial y \right)/B_0R_0$.  Recall that
this contribution is neglected in the density dynamics for $\alpha = 0$ in the upper row.
In both cases the flow field advects the blob radially outwards by transporting plasma
from the front of the blob along the equi-potential lines poloidally above and below its
density center into the wake of the blob. A finite electric drift compressibility inhibits
this transport along the poloidal flanks. This leads to a poloidal elongation of the blob,
as suggested in the lower left panel of the figure, and eventually to a dispersion of the
density into two poloidally separated structures.

\begin{figure}[htb]
    \includegraphics[width=\columnwidth]{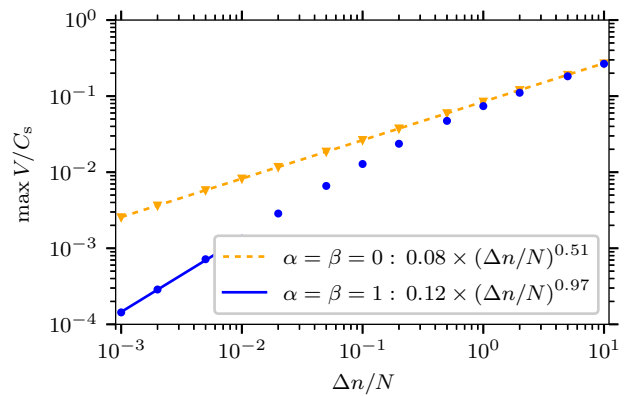}
    \caption{Maximal radial center of mass velocity as a function of initial amplitude 
             for $\kappa = 10^{-2}$. The dashed and full line indicate a least squares 
             fit to the maximal velocity for $\triangle n / N \leq 10^{-2}$.}
    \label{fig:vmax_deltanN}
\end{figure}

Figure \ref{fig:vmax_deltanN} presents the blobs maximal radial center of mass velocity
as a function of its initial perturbation amplitude. When drift compression is
absent, $\alpha = \beta = 0$, the maximal radial center of mass velocity of the blob 
is found to be proportional to the square root of its initial amplitude. This is in
agreement with \Eqnref{vorticity_scaling}. From a least squares fit of a power law
to the simulation data we evaluate $\mathcal{R} \approx 0.85$. This is in agreement
with previous studies of blob motion where a similar numerical value of $\mathcal{R}$
was found in the limit of negligible diffusion \Ref{garcia-2005, garcia-2006}.
On the other hand, for small amplitudes the blobs radial velocity depends linearly on its 
initial amplitude when incorporating drift compression terms in the density dynamics. A least 
squares fit of a power law for $\triangle n / N \leq 10^{-2}$ yields $\mathcal{P} \approx 0.50$. 

\begin{figure}[htb]
    \includegraphics[width=\columnwidth]{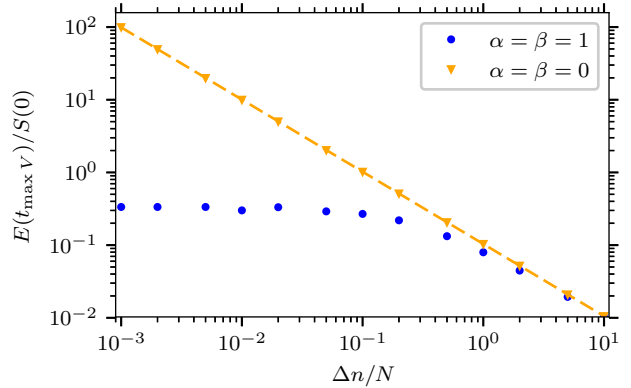}
    \caption{(Color) Ratio of the kinetic energy at the time the blob is propagating at 
             its maximal radial velocity to the initial value of $S$. The dashed line 
             indicates a least squares fit of a power law to the simulation data with an 
             exponent given by $-1$.}        
        \label{fig:energy_deltanN}
\end{figure}

Figure \ref{fig:energy_deltanN} shows the kinetic energy normalized to the initial
value of the entropy at the time $\tmax$ when the blob propagates at maximal radial velocity. 
Neglecting drift compression, maxima of this energy are up to two orders of magnitude larger 
than in simulations including drift compression for $\triangle n / N \ll 1$. For larger
initial amplitudes the kinetic energy of the blob when traveling at maximum center of mass 
velocity approaches values found in simulations of the model including drift compression. 
A power law fit to the simulation data suggests that the relative kinetic energy is inversely 
proportional to $\triangle n/N$. When electric drift compression is included in the model the 
kinetic energy is bounded by the initial free energy $S(0)$. For amplitudes 
$\triangle n/N \lesssim 0.5$ approximately one third of the initial free energy is converted 
to kinetic energy.

\begin{figure}[htb]
    \includegraphics[width=\columnwidth]{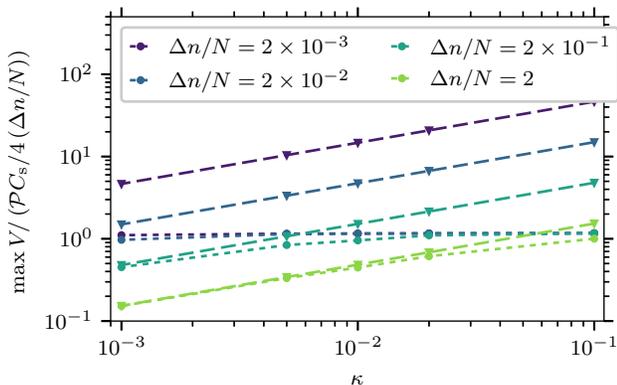}
    \caption{(Color) Maximal radial center of mass velocity relative to the scaling given 
             by \Eqnref{energy_scaling} as a function of the blob cross-field size 
             $\kappa$. The circles refer to simulation data of the compressional model, 
             $\alpha=\beta=1$, and the triangles marks simulation data with 
             $\alpha=\beta=0$.}
    \label{fig:vmax_kappa_deltan}
\end{figure}

Comparing the radial center of mass velocity as a function of the blobs cross-field size,
shown in \Figref{vmax_kappa_deltan} reveals the different size scaling of the velocity in
the compressible and incompressible cases. For the incompressible case, marked by the
triangles, the radial center of mass velocity shows a square
root dependence on the blobs cross-field size, as suggested by the inertial scaling of
\Eqnref{vorticity_scaling}. When compression is included, the velocity is independent of
the cross-field size for small initial amplitudes.  For sufficiently large amplitudes, the
square root scaling with blob size is observed also for the compressible case in
\Figref{vmax_kappa_deltan}.  In agreement with \Eqnref{transition}, the numerical
simulations of the compressible model show that the transition indeed depends on both the
cross-field size and the amplitude.

\begin{figure}[htb]
    \includegraphics[width=\columnwidth]{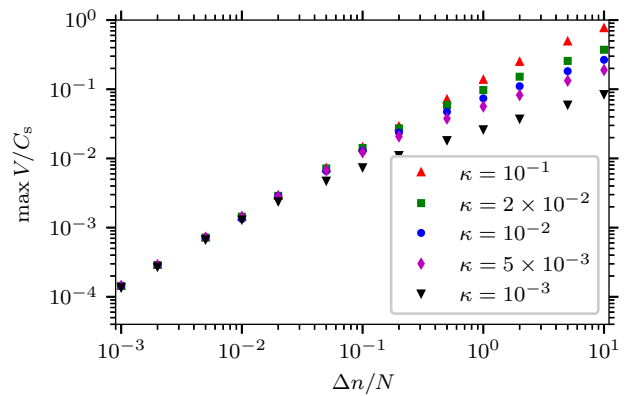}
    \caption{(Color) Maximal radial center of mass velocity as a function of its initial amplitude
             for varying ratios of the blob cross-field size to major radius.}
     \label{fig:vmax_deltan_kappa}
\end{figure}

The maximal radial center of mass velocities as a function of initial amplitude for
various cross-field sizes are shown in \Figref{vmax_deltan_kappa}. For small amplitudes,
$\triangle n / N \lesssim 10^{-2}$, the maximal radial velocity becomes independent of the
blobs cross-field size and depends approximately linearly on its initial amplitude. For
larger initial amplitudes the maximal velocity transitions into the square root dependence
on the blobs initial amplitude. Furthermore, $\max V$ depends on $\kappa$ for large
$\triangle n / N$ as also seen in \Figref{vmax_kappa_deltan}. This dependence is in
excellent agreement with the predictions from \Eqsref{vorticity_scaling} and
\ref{eq:energy_scaling}.

\begin{figure}[htb]
    \includegraphics[width=\columnwidth]{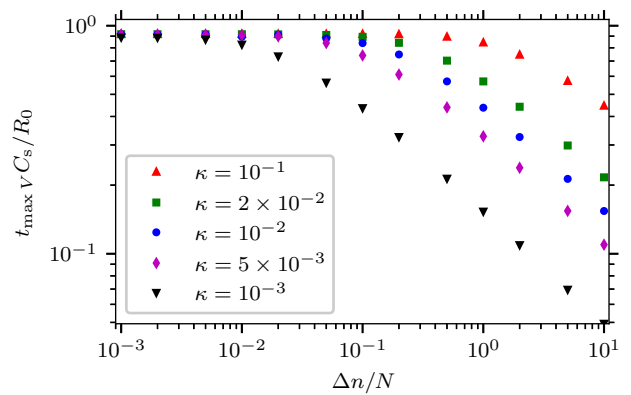}
    \caption{(Color) Time at which the blob propagates at maximal radial center of mass
             velocity as a function of its initial amplitude.}
    \label{fig:tmaxv_deltan_kappa}
\end{figure}

The numerical simulations further demonstrate that the time at which the blob assumes its
maximal radial center of mass velocity becomes independent of its initial amplitude in the
limit of small initial amplitudes, as shown in \Figref{tmaxv_deltan_kappa}. This is in
agreement with \Eqnref{tmaxv}. The range over which $\tmax$ is independent of $\triangle n
/ N$, is consistent with the critical initial amplitude \Eqnref{transition}.

\begin{figure}[htb]
    \includegraphics[width=\columnwidth]{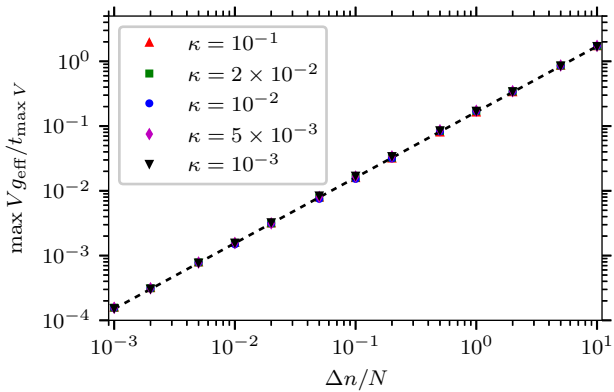}
    \caption{(Color) Maximal radial center of mass velocity of the blob divided by the 
             time at which the blob assumes this velocity. The dashed line denotes a 
             power law fit with exponent $1.0$ from which we evaluate
             $\mathcal{Q} \approx 0.34$.}
    \label{fig:acceleration}
\end{figure}

The simulations show furthermore that all blobs feature an initial period of constant
acceleration with the acceleration proportional to its initial amplitude and independent
of its cross-field size. This is clearly seen in \Figref{acceleration} and in agreement
with \Eqnref{acceleration}.  This corroborates the basis of the scale analysis leading to
\Eqnref{vorticity_scaling}, namely that initially the blob is subject to uniform
acceleration by an effective gravity, given by the interchange term in
\Eqnref{model_omega}.

The break in slopes for fixed $\kappa$ in the simulation data shown in
\Figsref{vmax_deltan_kappa} and \ref{fig:tmaxv_deltan_kappa} is predicted to appear at an
amplitude given by \Eqnref{transition}. Earlier in this section we obtained $\mathcal{P}
\approx 0.50$ by a least squares fit of \Eqnref{energy_scaling} to simulation data for the
compressible case with $\kappa = 10^{-2}$. A least squares fit of
\Eqnref{vorticity_scaling} to simulation data for the incompressible flow model 
gives $\mathcal{R} \approx 0.85$, such that $\deltantrans \approx 46 \kappa$. We
continue by comparing this predicted value to the simulation data.

\begin{figure}[htb]
    \includegraphics[width=\columnwidth]{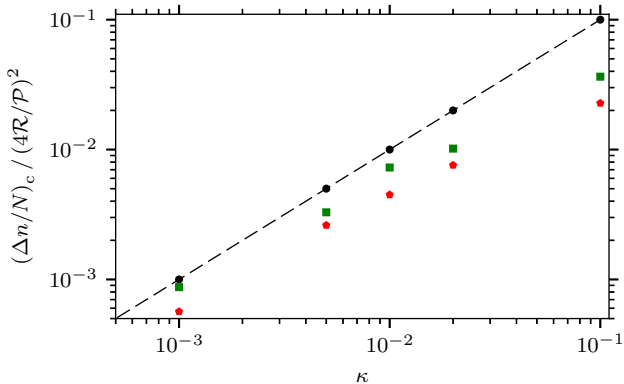}
    \caption{(Color) The critical transition amplitude \Eqnref{transition} evaluated with 
             $\mathcal{P} = 0.50$ and $\mathcal{R} = 0.85$ (black circles) compared
             to estimates of the transition amplitude from the velocity
             parameter scan (green squares) and the corresponding times at which 
             the blob propagates at its maximal radial velocity (red pentagon).}
    \label{fig:transition_estimate}
\end{figure}

To this end we fit \Eqsref{vorticity_scaling} and \ref{eq:energy_scaling} to the
simulation data shown in \Figref{vmax_deltan_kappa}, as well as \Eqnref{tmaxv} to the
simulation data shown in \Figref{tmaxv_deltan_kappa}. The fit range is chosen to be the
range where the data points approximately follow a power law scaling and the estimated
transition density amplitude is given by the intersection of the extrapolated fits.
\Figref{transition_estimate} compares the transition amplitude estimated by intersection
of fits to the velocity data (green squares) and fits to the $\tmax$ data (red pentagon)
to \Eqnref{transition} (black circles).

For small blob cross-field sizes, $\kappa = 10^{-3}$, the intersection of the fits to the
velocity data yields $\deltantrans \approx 4.0 \times 10^{-2}$,  the transition of the
fits to the $\tmax$ data yields $\deltantrans \approx  2.5 \times 10^{-2}$ while
\Eqnref{transition} yields $\deltantrans \approx 4.6 \times 10^{-2}$.  The discrepancy
between the estimates increases with $\kappa$ as the amplitude range over which the simulation data
follows an exact power law decreases. For $\kappa = 10^{-1}$ the intersection of the fits to
the $\max V$ data yields $\deltantrans \approx 1.7$, the intersection of fits to the
$\tmax$ data yields $\deltantrans \approx 1.0$ while \Eqnref{transition} evaluates to
approximately $4.6$. These values give a relative error of $0.61\ (0.78)$ on the
transition amplitude, given by \Eqnref{transition}, when compared to the $\max V$
($\tmax$) data.

\section{Discussion and Conclusions}
In conclusion, we have analyzed a two-field fluid model commonly used to describe blob
dynamics in the scrape-off layer of magnetically confined plasmas, in basic laboratory
plasma experiments and irregularities in ionospheres of celestial bodies. In section 2 we
discussed a commonly employed simplification, namely neglecting the compression of the
electric and diamagnetic drift in the particle continuity equation, and showed that
neglecting the electric drift compression corresponds to reducing the number of
conservation laws of the system.  In the compressible case the initialization of the
density field introduces an additional constraint on the energy conservation. This
introduces an upper limit on the blobs radial center of mass velocity for small amplitudes
and suggests that a velocity scaling regime exists where the blobs center of mass
velocity depends linearly on its initial perturbation amplitude and is independent on its
cross-field size, $V \sim \triangle n / N$. 

For this linear scaling, the time it takes a blob to assume its maximal radial velocity is
independent of its cross-field size or amplitude.  A scale analysis of the model equations
suggests on the other hand that the center  of mass velocity scales with the square root of
the blobs initial amplitude and cross-field size.  For this velocity scaling, $t_{\max V}$
is proportional to $\left( \ell / R_0 \right)^{1/2}$ as well as to 
$\left(\triangle n / N\right)^{-1/2}$. 
Assuming a smooth transition between these two velocity scalings in the compressible model 
we find that the transition point depends on both, the initial amplitude of the blob and 
its cross-field size. For the incompressible model, the cross-field size may be absorbed 
in the normalization of the model \Ref{garcia-2005} and only the square root velocity scaling 
is expected to hold.

In section 3 we presented numerical simulations of seeded blobs using the reduced model
equations. Both scaling regimes are recovered and the transition between them is found to
depend on the postulated parameters.
We estimated the numerical parameters for the velocity scaling laws given by
\Eqsref{vorticity_scaling} and \Eqnref{energy_scaling}, the transition amplitude 
given by \Eqnref{transition} and the time at which the blob propagates at its maximal
radial velocity given by \Eqnref{tmaxv} for a fixed initial blob cross-field size. The 
predicted transition point between the velocity scalings is shown to agree with the 
transition point found by fits to data from numerical simulations.

A large number of publications studying the dynamical properties of seeded blob
structures in tokamak scrape-off layers employ models in which compressional effects 
in the particle density dynamics are neglected 
\Ref{bian-2003, dippolito-2003, myra-2004, garcia-2005, myra-2006, garcia-2006, krash-2008, kube-2011, angus-2012}.
This is the first contribution to show that for typical blob cross-field sizes their 
dynamics is insensitive to electric drift compression only when the initial blob amplitude 
exceeds approximately half the background density. In this amplitude range the Boussinesq 
approximation used in the previous references and in this work is not strictly valid. 
Conclusively, we find that models incorporating an inhomogeneous magnetic field need 
to retain drift compression terms.

A direct application of velocity scaling laws for plasma filaments is to apply them to the
problem of broad particle density profiles observed in the scrape-off layer of
magnetically confined plasmas. A recently developed stochastic model predicts that the
density profile is proportional to the average blob amplitude and the duration time in
which a single blob traverses a given point 
\Ref{garcia-2016, garcia-2013, theodorsen-2016, garcia-2012, kube-2014}. 
On the other hand the duration time is given by the ratio of the blobs cross-field size to
its radial velocity. In turn, this allows to refine the dependence of the particle density
profile on blob properties and to test these predictions against experimental
measurements.

Future work will study the robustness of the derived scalings by comparing them to more
involved gyro-fluid simulations that relieve the Boussinesq approximation.  It is further
planned to elucidate the effect of energy conservation on blob dynamics in fluid models,
which parameterize the parallel dynamics and include the effects of magnetic field lines
intersecting material walls. 

\section{Acknowledgements}
We would like to acknowledge fruitful discussions with M.~Held.  The authors were
supported with financial subvention from the Research Council of Norway under grant
240510/F20. M.W. was supported by the Austrian Science Fund (FWF) Y398.  The computational
results presented have been achieved in part using the Vienna Scientific Cluster (VSC) and
by the Norwegian meta center for computational science (NOTUR) program through grants of
computation time



\begin{thebibliography}{1}
    \bibitem{zweben-1985} S.\ J.\ Zweben, Phys.\ Fluids {\bf 28} 974 (1985).
    \bibitem{antar-2001} G.\ Y.\ Antar, S.\ I.\ Krasheninnikov, P.\ Devynck, R.\ P.\ Doerner, E.\ M.\ Hollmann, 
                         J.\ A.\ Boedo, S.\ C.\ Luckhardt and R.\ W.\ Conn, Phys. Rev. Lett {\bf 87} 065001 (2001);
                         G.\ Y.\ Antar, P.\ Devynck, X.\ Garbet and S.\ C.\ Luckhardt, Phys. Plasmas {\bf 8} 1612 (2001);
                         G.\ Y.\ Antar, G.\ Counsell, Y.\ Yu, B.\ LaBombard and P.\ Devynck, Phys. Plasmas {\bf 10} 419 (2003) and references therein.
    \bibitem{boedo-2003} J.\ A.\ Boedo, D.\ L.\ Rudakov, R.\ J.\ Colchin, R.\ A.\ Moyer, S.\ Krasheninnikov, 
                         D.\ G.\ Whyte, G.\ R.\ McKee, G.\ Porter, M.\ J.\ Schaffer, 
                         P.\ C.\ Stangeby, W.\ P.\ West, S.\ L.\ Allen, and A.\ W.\ Leonard,
                         Journ.\ Nucl. Mater. {\bf 313-316} 813 (2003).
    \bibitem{terry-2003} J.\ L.\ Terry, S.\ J.\ Zweben, K.\ Hallatscheck, B.\ LaBombard,
                         R.\ J.\ Maqueda, B.\ Bai, C.\ J.\ Boswell, M.\ Greenwald, D.\ Kopon,
                         W.\ M.\ Nevins, C.\ S.\ Pitcher, B.\ N.\ Rogers, D.\ P.\ Stotler, and X.\ Q.\ Xu,
                         Phys.\ Plasmas {\bf 10} 1739 (2003).
    \bibitem{agostini-2007} M.\ Agostini, S.\ J.\ Zweben, R.\ Cavazzana, P.\ Scarin, G.\ Serianni,
                            R.\ J.\ Maqueda, and D.\ P.\ Stotler, Phys.\ Plasmas {\bf 14} 102305 (2007).
    \bibitem{garcia-2013} O.\ E.\ Garcia, S.\ M.\ Fritzner, R.\ Kube, I.\ Cziegler, B.\ LaBombard, and J.\ L.\ Terry,
                          Phys.\ Plasmas {\bf 20} 055901 (2013).
    \bibitem{umansky-1998} M.\ Umansky, S.\ I.\ Krasheninnikov, B.\ LaBombard and J.\ L.\ Terry, Phys.\ Plasmas {\bf 5} 3373 (1998).
    \bibitem{lipschultz-2001} B.\ Lipschultz, B.\ LaBombard, C.\ S.\ Pitcher, and R.\ Boivin, Plasma Phys. Control. Fusion {\bf 44} 733 (2002).
    \bibitem{labombard-2001} B.\ LaBombard, R.\ L.\ Boivin, M.\ Greenwald, J.\ Hughes, B.\ Lipschultz, 
                             D.\ Mossessian, C. S. Pitcher, J. L. Terry, S. J. Zweben,b)
                             and Alcator Group, Phys.\ Plasmas {\bf 8} 2107 (2001).
    \bibitem{carralero-2015} D.\ Carralero, P.\ Manz, L.\ Aho-Mantila, G.\ Birkenmeier, M.\ Brix, M.\ Groth, 
                             H.\ W.\ Müller, U.\ Stroth, N.\ Vianello, E.\ Wolfrum, 
                             ASDEX Upgrade team, JET Contributors, and EUROfusion MST1 Team, 
                             Phys.\ Rev.\ Letters {\bf 115} 215002 (2015);
                             D.\ Carralero, G.\ Birkenmeier, H.\ W.\ Müller, P.\ Manz, P.\ deMarne, S.\ H.\ Müller, 
                             F.\ Reimold, U.\ Stroth, M.\ Wischmeier, E.\ Wolfrum and The ASDEX Upgrade Team, 
                             Nucl.\ Fusion {\bf 54} 123005 (2014).
    \bibitem{kube-2016} R.\ Kube, A.\ Theodorsen, O.\ E.\ Garcia, B.\ LaBombard and J.\ L.\ Terry, 
                        Plasma Phys.\ Control. Fusion {\bf 58} 054001 (2016).
    \bibitem{garcia-2016} O.\ E.\ Garcia, R.\ Kube, A.\ Theodorsen, and H.\ L.\ Pécseli, Phys.\ Plasmas {\bf 23} 052308 (2016).
    \bibitem{dippolito-2002} D.\ A.\ D'Ippolito, J.\ R.\ Myra, and S.\ I.\ Krasheninnikov, Phys.\ Plasmas {\bf 9} 222 (2002).
    \bibitem{bian-2003} N.\ Bian, S.\ Benkadda, J.\ V.\ Paulsen, and O.\ E.\ Garcia, Phys.\ Plasmas {\bf 10} 671 (2003).
    \bibitem{garcia-2005} O.~E.~Garcia, N.~H.~Bian, V.~Naulin, A.~H.~Nielsen, and J.~Juul Rasmussen, Phys.\ Plasmas {\bf 12} 090701 (2005).
    \bibitem{krash-2001} S.\ I.\ Krasheninnikov, Phys.\ Lett.\ A {\bf 283} 368 (2001).
    \bibitem{dippolito-2004} D.\ A.\ D’Ippolito, J.\ R.\ Myra, S.\ I.\ Krasheninnikov, G.\ Q.\ Yu, and 
                             A.\ Yu. Pigarov, Contrib.\ Plasma Phys.\ {\bf 44} 205 (2004).
    \bibitem{dippolito-2011} D.\ A.\ D'Ippolito, J.\ R.\ Myra and S.\ J.\ Zweben, Phys.\ Plasmas {\bf 18} 060501 (2011).
    \bibitem{garcia-2006} O.~E.~Garcia, N.~H.~Bian, W.~Fundamenski, Phys.\ Plasmas {\bf 13} 082309 (2006).
    \bibitem{garcia-2006-ps} O.\ E.\ Garcia, N.\ H.\ Bian, V.\ Naulin, A.\ H.\ Nielsen and J.\ Juul Rasmussen, Phys.\ Scr.\ {\bf T122} 104 (2006).
    \bibitem{rudakov-2005} D.\ L.\ Rudakov, J.\ A.\ Boedo, R.\ A.\ Moyer, P.\ C.\ Stangeby, 
                           J.\ G.\ Watkins, D.\ G.\ Whyte, L.\ Zeng, N.\ H.\ Brooks,
                           R.\ P.\ Doerner, T.\ E.\ Evans, M.\ E.\ Fenstermacher, M.\ Groth, 
                           E.\ M.\ Hollmann, S.\ I.\ Krasheninnikov, C.\ J.\ Lasnier,
                           A.\ W.\ Leonard, M.\ A.\ Mahdav, G.\ R.\ McKee, A.\ G.\ McLean, 
                           A.\ Yu.\ Pigarov, W.\ R.\ Wampler, G.\ Wang, W.\ P.\ West and
                           C.\ P.\ C.\ Wong, Nucl.\ Fusion {\bf 45} 1589 (2005).
    \bibitem{grulke-2006} O.\ Grulke, J.\ L.\ Terry, B.\ Labombard, and S.\ J.\ Zweben, Phys.\ Plasmas {\bf 13} 012306 (2006).
    \bibitem{grulke-2014} O.\ Grulke, J.\ L.\ Terry, I.\ Cziegler, B.\ LaBombard and O.\ E.\ Garcia, Nucl.\ Fusion {\bf 54} 043012 (2014).
    \bibitem{theodorsen-2016} A.\ Theodorsen, O.\ E.\ Garcia, J.\ Horacek, R.\ Kube, and R.\ A.\ Pitts, Plasma Phys.\ Control.\ Fusion {\bf 58} 044006 (2016).
    \bibitem{katz-2008} N.~Katz, J.~Egedal, W.~Fox, A.~Le, and M.~Porkolab, Phys.\ Rev.\ Lett.\ {\bf 101} 015003 (2008).
    \bibitem{theiler-2009} C.~Theiler, I.~Furno, P.~Ricci, A.~Fasoli, B.~Labit, S.H.~Müller, and G.~Plyushchev, Phys.\ Rev.\ Lett.\ {\bf 103} (2009).
    \bibitem{furno-2011} I.\ Furno, C.\ Theiler, D.\ Lançon, A.\ Fasoli, D.\ Iraji, P.\ Ricci, M.\ Spolaore and N.\ Vianello, Plasma Phys. Control. Fusion {\bf 53} 124016 (2011).
    \bibitem{riva-2016} F.\ Riva, C.\ Colin, J.\ Denis, L.\ Easy, I.\ Furno, J.\ Madsen, F.\ Militello, V.\ Naulin, A.\ H.\ Nielsen, J.\ M.\ B.\ Olsen, J.\ T.\ Omotani, J.\ Juul Rasmussen, P.\ Ricci, E.\ Serre, P.\ Tamain and C.\ Theiler, Plasma Phys. Control. Fusion {\bf 58} 044005 (2016).
    \bibitem{hysell-2000} D.\ L.\ Hysell, Journ.\ Atmos.\ Solar-Terr.\ Phys.\ {\bf 62} 1037 (2000).
    \bibitem{rymer-2009} A.\ M.\ Rymer, B.\ H.\ Mauk, T.\ W.\ Hill, N.\ Andre, D.\ G.\ Mitchell, C.\ Paranicas, D.\ T.\ Young, H.\ T.\ Smith, A.\ M.\ Persoon, J.\ D.\ Menietti, G.\ B.\ Hospodarsky, A.\ J.\ Coates, M.\ K.\ Dougherty, Planet.\ Space Science {\bf 57} 1779 (2009).
    \bibitem{wolf-2012} R.\ A.\ Wolf, C.\ X.\ Chen, and F.\ R.\ Toffoletto, Journ. Geophys. Res. {\bf 117} A02215 (2012).
    \bibitem{dippolito-2003} D.\ A.\ D’Ippolito and J.\ R.\ Myra, Phys. Plasmas {\bf 10} 4029 (2003)
    \bibitem{myra-2004} J.\ R.\ Myra, D.\ A.\ D'Ippolito, S.\ I.\ Krasheninnikov and G.\ Q.\ Yu, Phys.\ Plasmas {\bf 11} 4267 (2004).
    \bibitem{yu-2006} G.\ Q.\ Yu, S.\ I.\ Krasheninnikov, and P.\ N.\ Guzdar, Phys. Plasmas {\bf 13} 042508 (2006).
    \bibitem{angus-2014} J. Angus, and S. I. Krasheninnikov, Phys.\ Plasmas {\bf 21} 112504 (2014)
    \bibitem{omotani-2015} J.\ T.\ Omotani, F.\ Militello, L.\ Easy and N.\ R.\ Walkden, Plasma Phys.\ Control Fusion {\bf 58} 014030 (2015).
    \bibitem{held-2016} M. Held, M. Wiesenberger, J. Madsen, and A. Kendl, Nucl.\ Fusion {\bf 56} 126005 (2016).
    \bibitem{wiesenberger-2014} M.~Wiesenberger, J.~Madsen, and A.~Kendl, Phys.\ Plasmas {\bf 21} 092301 (2014).
    \bibitem{madsen-2011} J.\ Madsen, O.\ E.\ Garcia, J.\ S.\ Larsen, V.\ Naulin, A.\ H.\ Nielsen, and J.\ Juul Rasmussen, Phys.\ Plasmas {\bf 18} 112504 (2011).
    \bibitem{manz-2013} P.\ Manz, D.\ Carralero, G.\ Birkenmeier, H.\ W.\ Müller, S.\ H.\ Müller, G.\ Fuchert, B.\ D.\ Scott, and U.\ Stroth, Phys.\ Plasmas {\bf 20} 102307 (2013)
    \bibitem{kube-2011} R.\ Kube and O.\ E.\ Garcia, Phys.\ Plasmas {\bf 18} 102314
                        (2011);
                        R.\ Kube and O.\ E.\ Garcia, Phys.\ Plasmas {\bf 19} 042305
                        (2012).
    \bibitem{myra-2005} J.~R.~Myra and D.~I.~D'Ippolito, Phys.\ Plasmas {\bf 12} 092511 (2005).
    \bibitem{olsen-2016} J.\ Olsen, J.\ Madsen, A.\ H.\ Nielsen, J.\ Juul Rasmussen and V.\ Naulin, Plasma Phys.\ Control.\ Fusion {\bf 58} 044011 (2015).
    \bibitem{ott-1978} E.\ Ott, J.\ Geophys.\ Res. {\bf 83}, 2066, (1978)
    \bibitem{scott-2005} B.\ Scott, Phys.\ Plasmas {\bf 12} 102307 (2005).
    \bibitem{2dads} \url{https://github.com/rkube/2dads}
    \bibitem{wiesenberger-phd} M.~Wiesenberger, Gyrofluid computations of filament dynamics in tokamak scrape-off layers, 
        \href{http://diglib.uibk.ac.at/urn:nbn:at:at-ubi:1-1799}{PhD thesis} (2014), University of Innsbruck, Austria; 
        \url{https://github.com/feltor-dev/feltor}
    \bibitem{numerics-supplemental} Replication Data for: Amplitude scaling for interchange motions of plasma filaments  \href{http://dx.doi.org/10.18710/7RRESR}{http://dx.doi.org/10.18710/7RRESR}
    \bibitem{muller-2009} S.~H.~Müller, C.~Theiler, A.~Fasoli, I.~Furno, B.~Labit, G.~R.~ Tynan, M.~Xu, Z.~Yan and J.~H.~Yu, Plasma Phys.\ Control. Fusion {\bf 51} 055020 (2009).
    \bibitem{angus-2012} J.\ R.\ Angus, S.\ I.\ Krasheninnikov, and M. V.\ Umansky, Phys.\ Plasmas {\bf 19} 082312 (2012).
    \bibitem{krash-2008} S.\ I.\ Krasheninnikov and S.\ I.\ Smolyakov, Phys.\ Plasmas {\bf 15} 055909 (2008).
    \bibitem{myra-2006} J.\ R.\ Myra, D.\ A. Russel, and D.\ A.\ D'Ippolito, Phys.\ Plasmas {\bf 13} 112502 (2006).
    \bibitem{garcia-2012} O.\ E.\ Garcia, Phys.\ Rev.\ Letters {\bf 108} 265001 (2012).
    \bibitem{kube-2014} R.\ Kube and O.\ E.\ Garcia, Phys.\ Plasmas {\bf 22} 012502 (2014). 
\end{thebibliography}
\end{document}